\documentclass[referee]{raa}            

\usepackage{graphicx,times}             
\usepackage{natbib}
\usepackage{amssymb,amsmath}
\usepackage{booktabs}
\bibpunct{(}{)}{;}{a}{}{,}

\usepackage[a4paper=true,pagebackref=true]{hyperref}
\hypersetup{colorlinks = true, linkcolor = green, anchorcolor = red, citecolor = blue, filecolor = red, pagecolor = red, urlcolor = red}

\begin{document}

   \title{Performance evaluation of baseline-dependent averaging based on full-scale SKA1-LOW simulation
}

   \volnopage{Vol.0 (20xx) No.0, 000--000}      
   \setcounter{page}{1}          

   \author{Qing-Wen Deng
      \inst{1,2}
   \and Feng Wang
      \inst{1,2}
   \and Hui Deng
      \inst{1,2}
   \and Ying Mei
      \inst{1,2}
   \and Jing Li
      \inst{1,2}      
   \and Oleg Smirnov
      \inst{3}
   \and Shao-Guang Guo\inst{4}
   }

   \institute{Center for Astrophscis, School of Physics and Materials Science, Guangzhou University, Guangzhou 51006, China; {\it fengwang@gzhu.edu.cn}\\
        \and
{Great Bay Center, National Astronomical Data Center, Guangzhou, Guangdong, China, 510006} \\
\and
{Department of Physics and Electronics, Rhodes University, PO Box 94, Makhanda, 6140, South Africa} \\
\and{Shanghai Astronomical Observatory, Chinese Academy of Sciences, Shanghai, China, 200030 }
\vs\no
   {\small Received~~2021 November 22; accepted~~2022~~February 16}}

\abstract{ The Square Kilometre Array (SKA) is the largest radio interferometer under construction in the world. Its immense amount of visibility data poses a considerable challenge to the subsequent processing by the science data processor (SDP). Baseline dependent averaging (BDA), which reduces the amount of visibility data based on the baseline distribution of the radio interferometer, has become a focus of SKA SDP development. This paper developed and implemented a full-featured BDA module based on Radio Astronomy Simulation, Calibration and Imaging Library (RASCIL). Simulated observations were then performed with RASCIL based on a full-scale SKA1-LOW configuration. The performance of the BDA was systematically investigated and evaluated based on the simulated data. The experimental results presented that the amount of visibility data is reduced by about 50\% to 85\% for different time intervals ($\Delta t_{max}$). In addition, different $\Delta t_{max}$ have a significant effect on the imaging quality. The smaller the $\Delta t_{max}$, the smaller the degradation of the imaging quality.
}

  \authorrunning{Q.-W. Deng, F. Wang, H. Deng, Y. Mei, J. Li, O. Smirnov \& S.-G. Guo }  
  \titlerunning{Performance evaluation of baseline-dependent averaging for SKA1-LOW}  

   \maketitle

%
%

\section{Introduction}           

The Square Kilometre Array (SKA) (\citealt{Braun+1996}) is an ongoing international project to build the world's largest radio interferometric telescope with more than one square kilometer potential collection area. With detailed design and preparation well underway, the SKA represents a giant leap forward in engineering to deliver a unique instrument.

The $uv$ distribution of a radio interferometer generally has a dense center and a sparse edge. With the rotation of the Earth, each sampling point of the radio interferometer draws an arc-shaped trajectory around the phase center on the $uv$ plane. Short baselines have denser data than long baselines for the same track length, and the difference can be even greater for a larger array.
Decorrelation can be avoided on the longer baselines when more samples are averaged at the center than at the outer edges. At the same time, data compression can be carried out on shorter baselines.

The baseline dependent averaging (BDA) was presented by \cite{Cotton+1986,Cotton+1999} to reduce the visibility data volumes, which has been used by the MWA (\citealt{Mitchell+2008}) and is also used to shape the field of interest (\citealt{Atemkeng+2018}).
However, averaging visibilities over time and frequency will cause image distortion, also called the smearing effect. The bandwidth smearing is manifested as a position-dependent and radial convolution effect in the image field. Time smearing is similar to bandwidth smearing but more complicated, described as a loss in amplitude (\citealt{Cotton+1986,Cotton+1999}). 
\cite{Bridle+1999} conducted a mathematical analysis of these two smearing effects and found that they cannot be effectively corrected by calibration or self-calibration methods. It is recommended to design a comprehensive observation strategy to reduce the impact to an acceptable level. Therefore, the short integration time and small channel width are necessary for the long baselines for a radio interferometer to suppress the smearing effect. In contrast, the resolution requirement for time and frequency is relatively low on the short baselines.

The BDA does not change the channel bandwidth and integration time for long baseline visibility data, but averaging of short baseline visibility data corresponds to an increase in integration time and channel bandwidth.
\cite{Wijnholds+2018} obtained the Cramer-Rao Bound of averaged visibilities by estimating the number of the raw visibilities and compared it with the covariance obtained by the error transfer formula. 
It is proven that BDA will not cause other effects except for the approximately obtainable decorrelation loss. \cite{Salvini+2017} proposed the Compress-Expand-Compress method to expand the visibilities to the required time resolution for calibration after the first compression, and then perform the second compression, and finally achieve a high compression ratio of the visibilities in time. However, few previous literature presented the quantitative analysis of BDA on the final image quality and storage costs.

As the SKA enters its construction phase and the SKA Science Regional Centers (SRCs) construction white paper has been released, it becomes imperative to research the BDA technique further and analyze its usability for the SKA1 scale. We wish to analyze and discuss this study systematically: 1. How much space would be saved by using BDA technology for SKA1-LOW observations? 2. Is there a significant degradation of the final dirty image with the BDA?

In the rest of this study, we first introduce the BDA algorithm and its implementation. We then simulate full-scale SKA1-LOW observations and investigate the BDA performance in Section 3. The discussions are described in Section 4. The conclusions and future work are presented in the last section. 

\section{The algorithm and implementation of BDA}
\subsection{The BDA Algorithm}
For a radio interferometer, a visibility function is obtained by correlating the signal collected by two antennas of each baseline with the same time interval $\delta t$ and frequency sampling interval $\delta f$. According to the mathematical definition of BDA~\citep{Wijnholds+2018}, we can average the raw visibilities and thus obtain the averaged visibilities. 
In data processing, from $P$ receiving antennas $P^2$ correlations are assumed to be collected in $K$ short-term integrations, either over time, frequency, or both. The raw visibility data vector can be defined as 
\begin{equation}
    \mathbf{r} = \left[ r_1, \cdots, r_K \right]^T \in \mathbb{C}^{KP^2 \times 1}
\end{equation}
where $\mathbb{C}$ denotes the complex matrix. And the averaging process can be described as
\begin{equation}
    \mathbf{r}_\mathrm{ave} = 
    \mathbf{W} \mathbf{I}_\mathrm{s} \mathbf{r} 
    \in \mathbb{C} ^ {M \times 1}
\end{equation}
where $\mathbf{W} = \left( \mathbf{I}_\mathrm{s} \mathbf{I}_\mathrm{s}^{T} \right) ^ {-1}$ is a diagonal weighting matrix. $\mathbf{I}_\mathrm{s} \in \mathbb{R}^{M \times KP^2}$ is a sparse matrix that determines which samples will be averaged together, where $\mathbb{R}$ denotes the real matrix and $M$ is the length of the average visibility data vector. 

Suppose we ignore the correlation effects and assume that the values of the visibility data averaged together are the same. In that case, the raw visibilities can be obtained approximately from the averaged visibilities by
\begin{equation}
    \mathbf{r} = \mathbf{I}_{\mathrm{s}}^T \mathbf{r}_{\mathrm{ave}}
\end{equation}

The selection matrix $\mathbf{I}_\mathrm{s}$ is related to the averaging intervals of time and frequency in the BDA. For a selected baseline $D$, the averaging intervals can be calculated by the rounding ratio of that baseline to the longest baseline $B_{max}$ as 
\begin{equation}
    t = \delta t \frac{B_{max}}{D},~~~~
    f = \delta f \frac{B_{max}}{D}
    \label{ms2021-0369equ4}
\end{equation}

We used the rounding ratio of $B_{max}$ to $D$ to determine whether a BDA processing is needed. On partial long baselines, when $\frac{B_{max}}{D} = 1$, the data will not be averaged. While on the short baselines, $\frac{B_{max}}{D} > 1$, indicating that more sampling data can be averaged. In addition, $t$ is limited by the calibration time scale determined by the environment and instrument for the interferometer. A larger averaging scale will also make the smearing effects more serious. Therefore, it is necessary to set reasonable upper limits for $t$ and $f$ in the implementation, respectively.

\subsection{Implementation}

We implemented a full-featured BDA module based on the 
Radio Astronomy Simulation, Calibration and Imaging Library (RASCIL)~\footnote{https://gitlab.com/ska-telescope/external/rascil}. The RASCIL is a pure Python software package suite for radio interferometer calibration and imaging algorithms, especially for SKA data processing.  
Since the public release of RASCIL, it has been widely used in data processing for some radio interference telescopes(e.g., \cite{Wei+2021} and so on). 
We developed a BDA module based on RASCIL, which has been released online \footnote{https://github.com/astronomical-data-processing/ska-bda}.

\begin{figure}[h]
    \centering
    \includegraphics[width=9.5cm, angle=0]{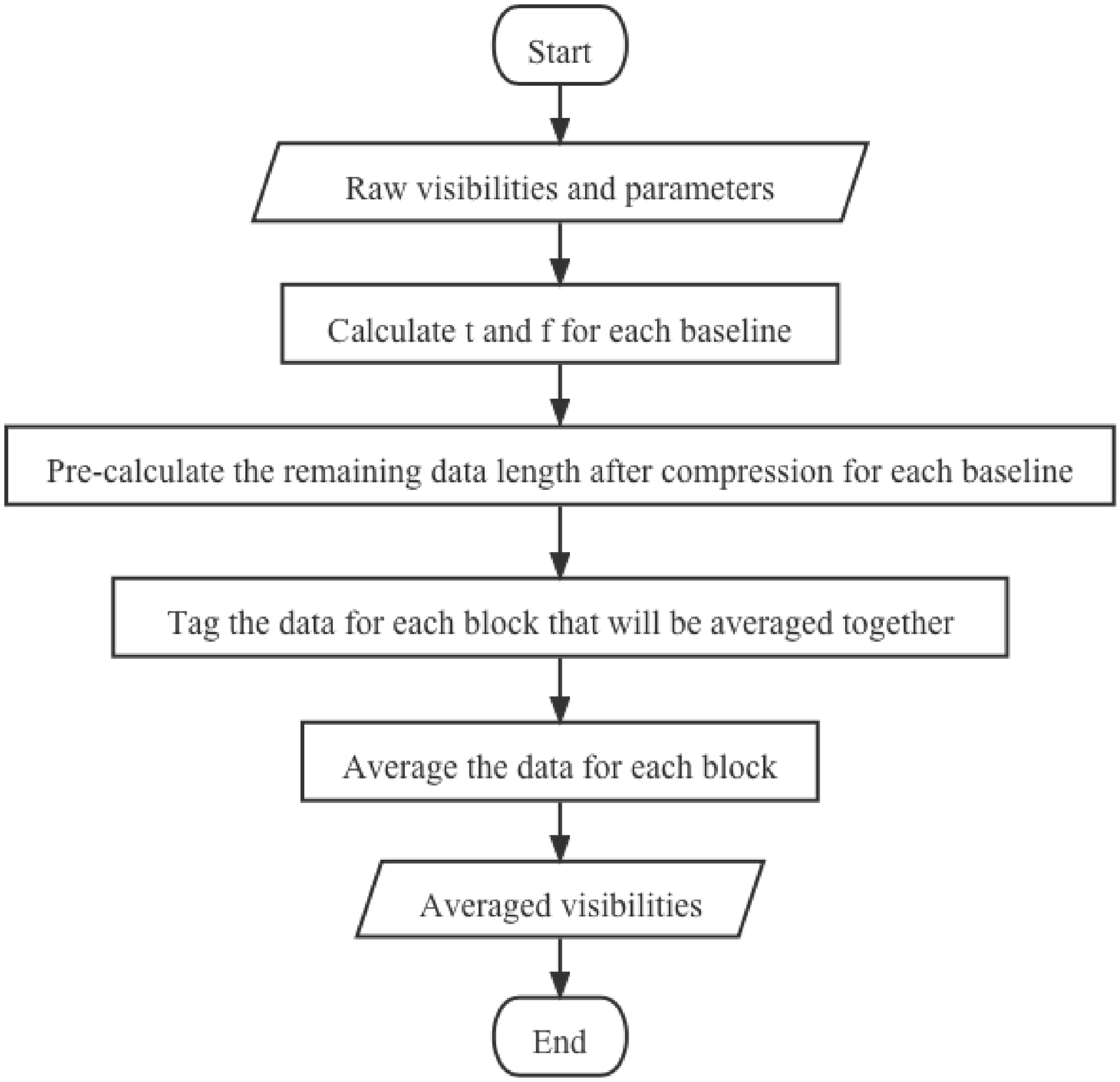}
    \caption{The flow chart of the BDA implementation.}
    \label{ms2021-0369fig1}
\end{figure}

\begin{figure}[h]
    \centering
    \includegraphics[width=9.1cm, angle=0]{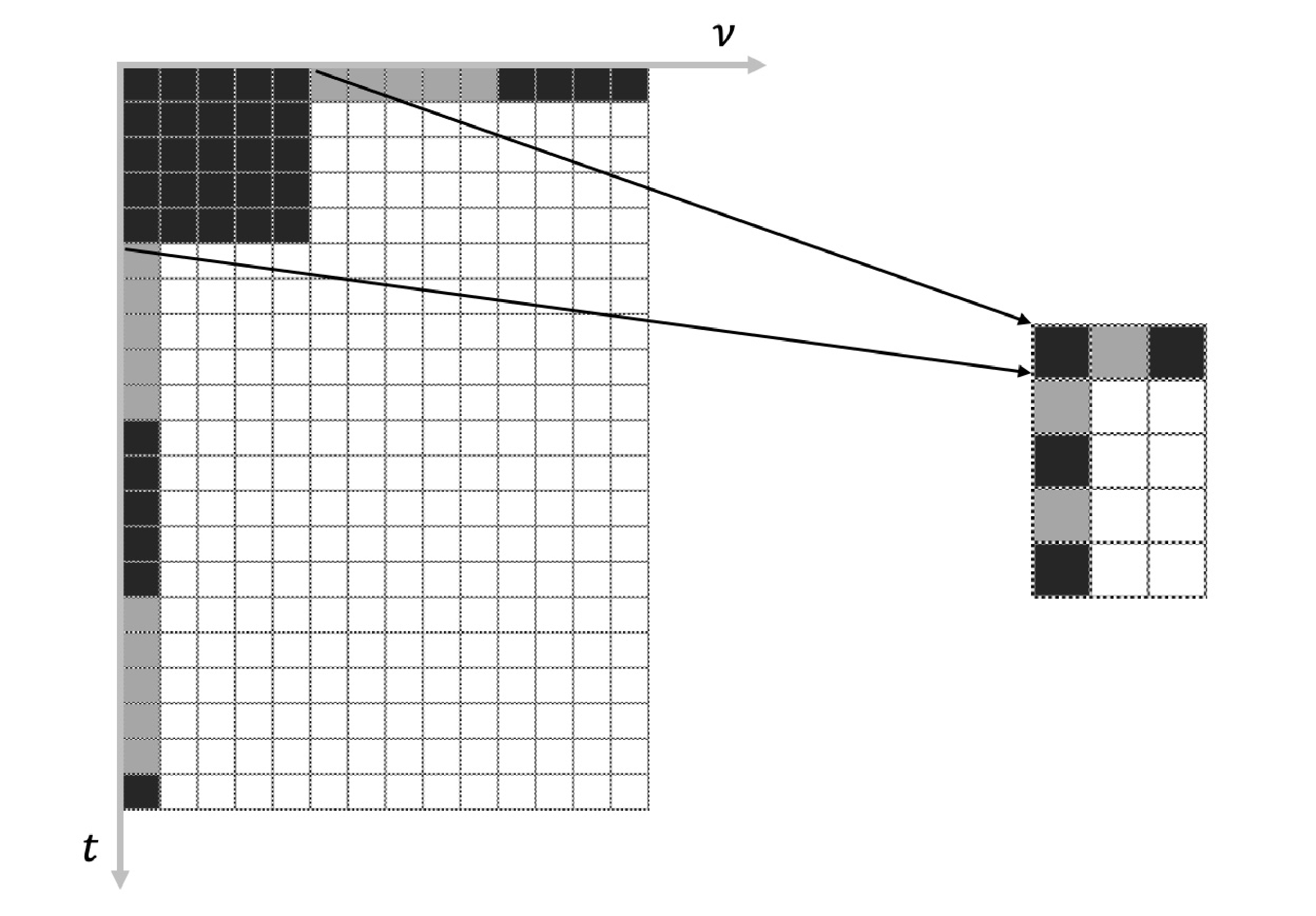}
    \caption{The averaging of visibility data at a baseline when applying BDA.}
    \label{ms2021-0369fig2}
\end{figure}

The flowchart of the implementation is shown in Figure \ref{ms2021-0369fig1}.
Figure \ref{ms2021-0369fig2} shows an example of averaging process on a baseline, where the data will be reduced from the original $21\times 14$ to $5\times 3$, by assuming that both $t$ and $f$ are 5 according to the Equation \ref{ms2021-0369equ4}. In the case of the shortest baselines, $t$ and $f$ are also equal to the upper limits of the time-frequency interval used in the BDA, defined as $\Delta t_{max}$ and $\Delta f_{max}$.
In the averaging process, we first calculate the position of the raw visibilities corresponding to the flattened averaged visibilities. We then average the visibilities based on the positions and the number of visibilities.

We used the BlockVisibility class defined in the RASCIL. The visibility data was stored using a multi-dimensional array, with dimensions including baseline, polarization, time and frequency.
To meet the requirements of BDA performance profiling, we implemented BDA by using three underlying packages, i.e., pure Python, Pandas, and Numba, respectively. 

In the pure Python implementation, we grouped the data for averaging based on Numpy. To optimize the performance, we tried to use Pandas, put all the parameter data into a table when preprocessing, and then performed group-by operations to complete all the calculations.

We also used Numba~(\citealt{Lam+2015}) to speed up the function and further improved the processing performance. Numba is an on-the-fly compiler that translates a subset of the Python and Numpy code used in the function into efficient machine code, which can effectively improve the speed of the program.

\section{Performance assessments For BDA}
\subsection{Observational configuration}

\begin{figure}[h]
    \centering
    \includegraphics[width=7cm, angle=0]{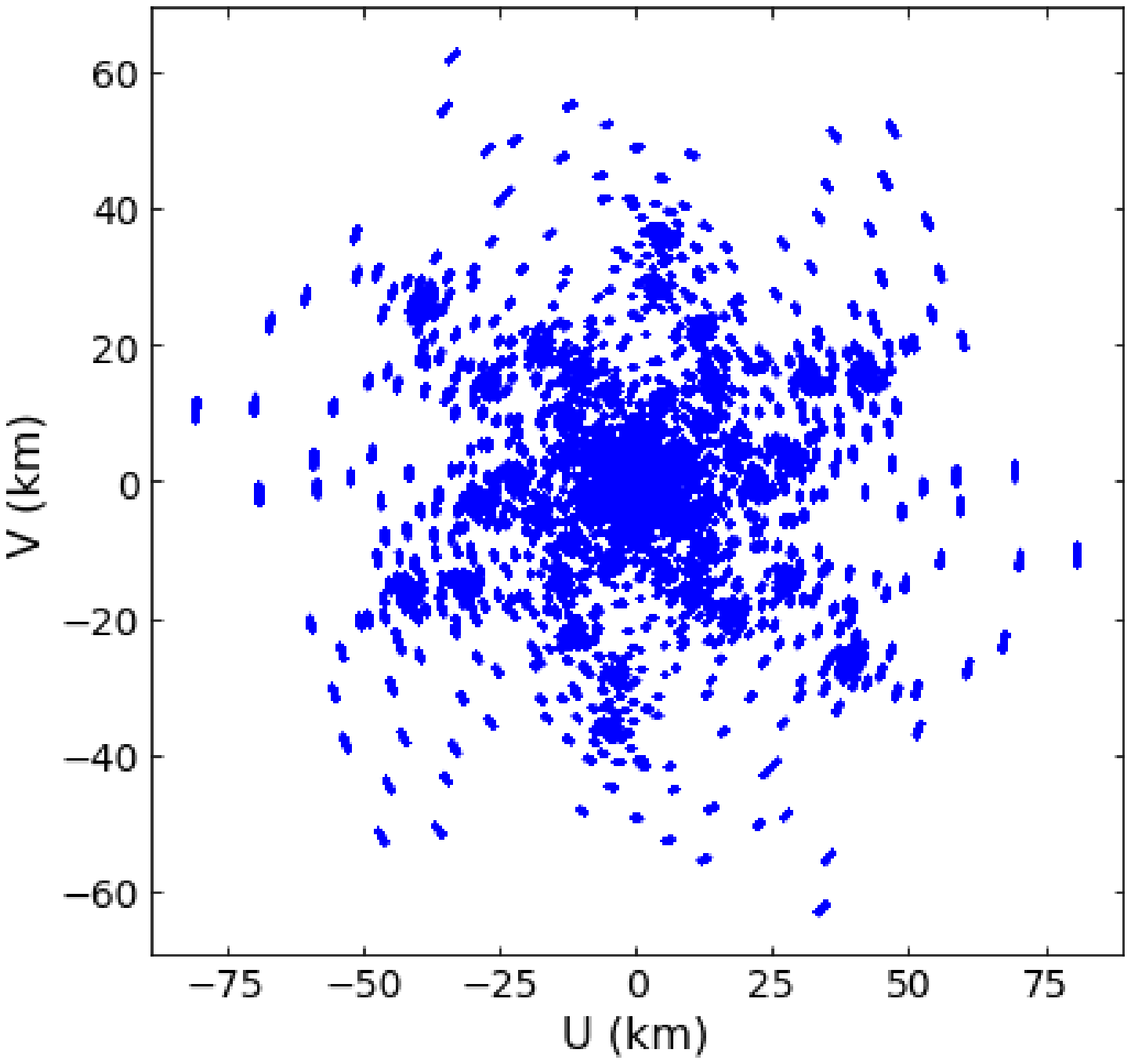}
    \caption{The $uv$ distribution of the SKA1-LOW observed in a single channel at 100 MHz for 12 minutes.}
    \label{ms2021-0369fig3}
\end{figure}

To more accurately evaluate the performance of the BDA, we used RASCIL to simulate single-channel and one polarization visibility data observed by the full scale SKA1-LOW telescope. During simulation, we used all SKA1-LOW 512 stations and set up 12 minutes of observations, of which 6 minutes on each side of the zenith. The integration time is set to 0.9 seconds as required by the array structure. The observing frequency is 100 MHz with the channel bandwidth 1 MHz, and the phase center in the observation points to the right ascension $15^{\circ}$ and the declination $-45^{\circ}$. We finally obtained 800 temporal sampling points on each baseline. The $uv$ distribution of simulated observation is shown in Figure \ref{ms2021-0369fig3}.

\subsection{Observation Simulation}

\begin{figure}[h]
    \begin{minipage}[t]{0.47\linewidth}
        \centering
        \includegraphics[width=6.6cm]{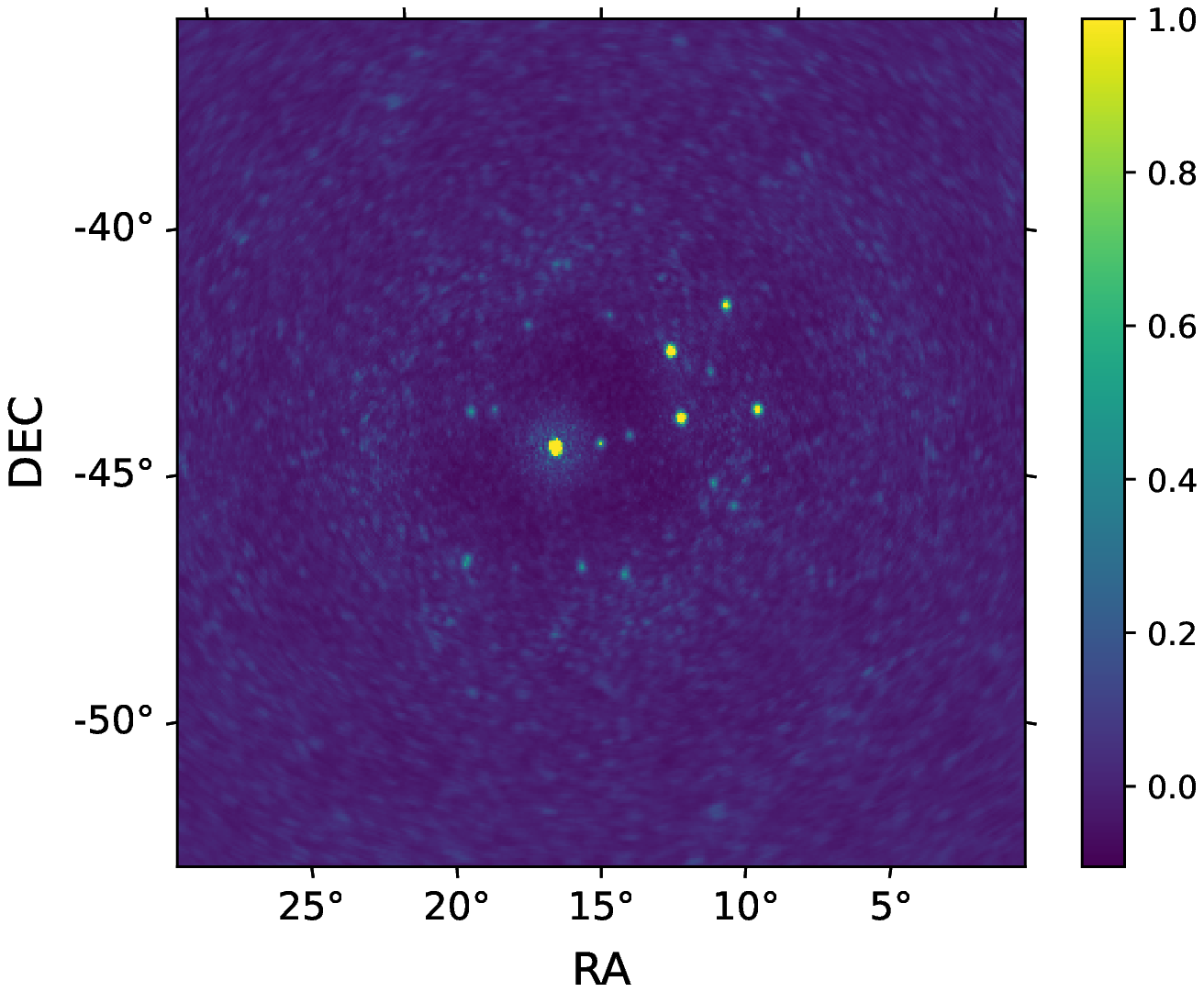}
    \end{minipage}%
    \begin{minipage}[t]{0.47\textwidth}
        \centering
        \includegraphics[width=7cm]{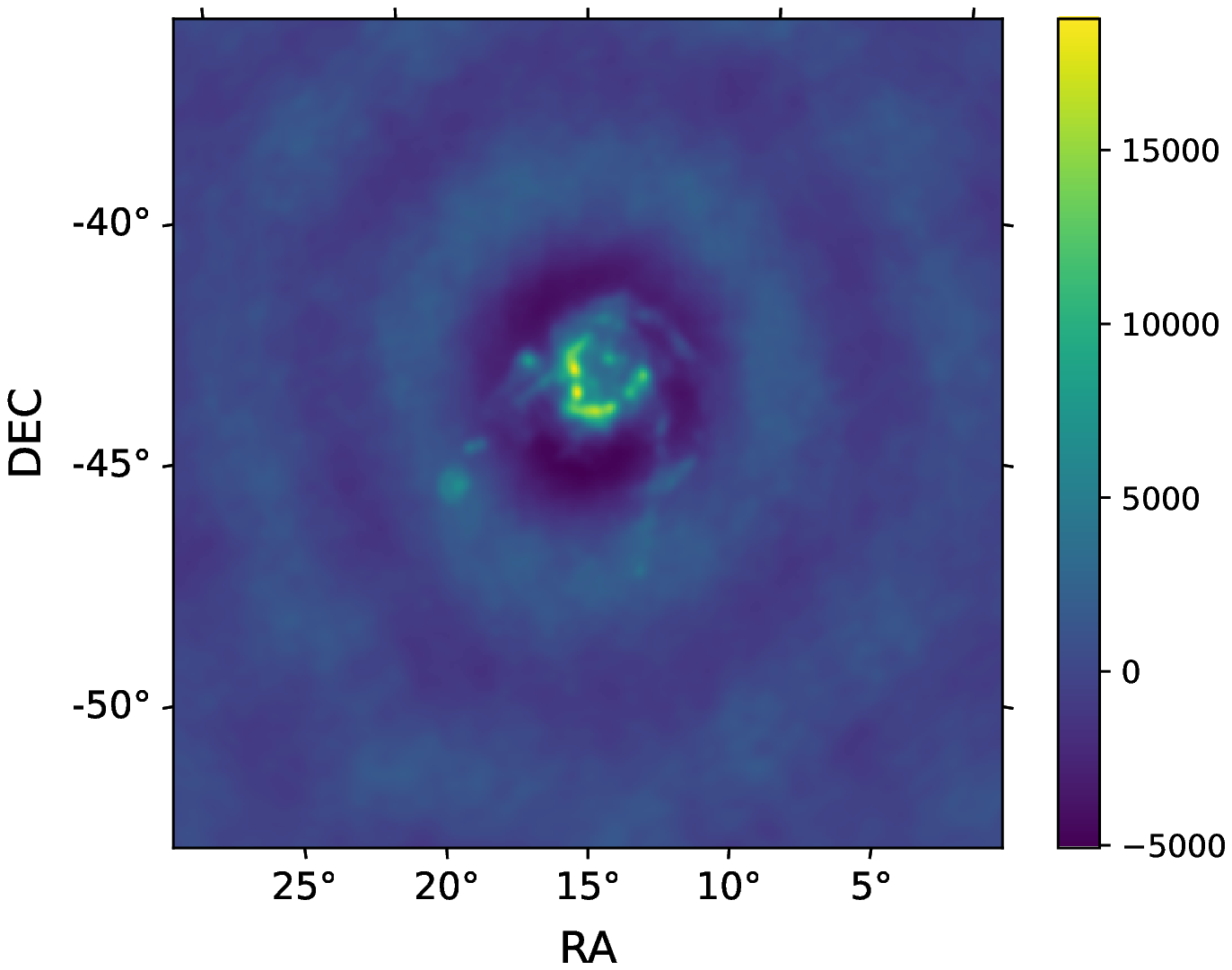}
    \end{minipage}%
    \centering
    \caption{Dirty images of the raw visibilities observed from the GLEAM model and the M31 model images.}
    \label{ms2021-0369fig4}
\end{figure}

With the observational configuration described above, we simulated observations of point and extend sources separately. We selected the corresponding sources from the GaLactic and Extragalactic All-sky Murchison Widefield Array (GLEAM) survey catalog \citep{GLEAM} with an imaging size of 32,768 multiplied by 32,768 pixel. 
The other is the M31 image that observed by Very Large Array. The image has a pixel size of $512\times512$ and a resolution of 1 arc second.  
We used the transform.resize() function in skimage \citep{Skimage+2014} to scale the M31 image to the same pixel scale as the image generated by GLEAM model for this study. It should be clear here that such a magnification of the original image is only necessary for the simulation of the observation.
Two dirty images for the cases are shown in Figure \ref{ms2021-0369fig4}.

\subsection{Evaluation Results}

We invoke the BDA module to perform visibilities compression, decompression, and imaging processing. Since BDA processing at frequency series has similar results at time series, we only simulate and analyze temporal BDA in this study. Also, the effects on visibility data and dirty images are investigated by setting different upper limits on time integration in the BDA.

\subsubsection{The compression ratio}

The compression ratio (CR) is calculated by comparing the data volumes between the averaged and raw visibilities, as $\mathrm{CR} = \frac{\mathbf{r}_\mathrm{ave}}{\mathbf{r}} \times 100\%$. 

\begin{figure}[h]
    \begin{minipage}[t]{0.47\linewidth}
        \centering
        \includegraphics[width=68mm]{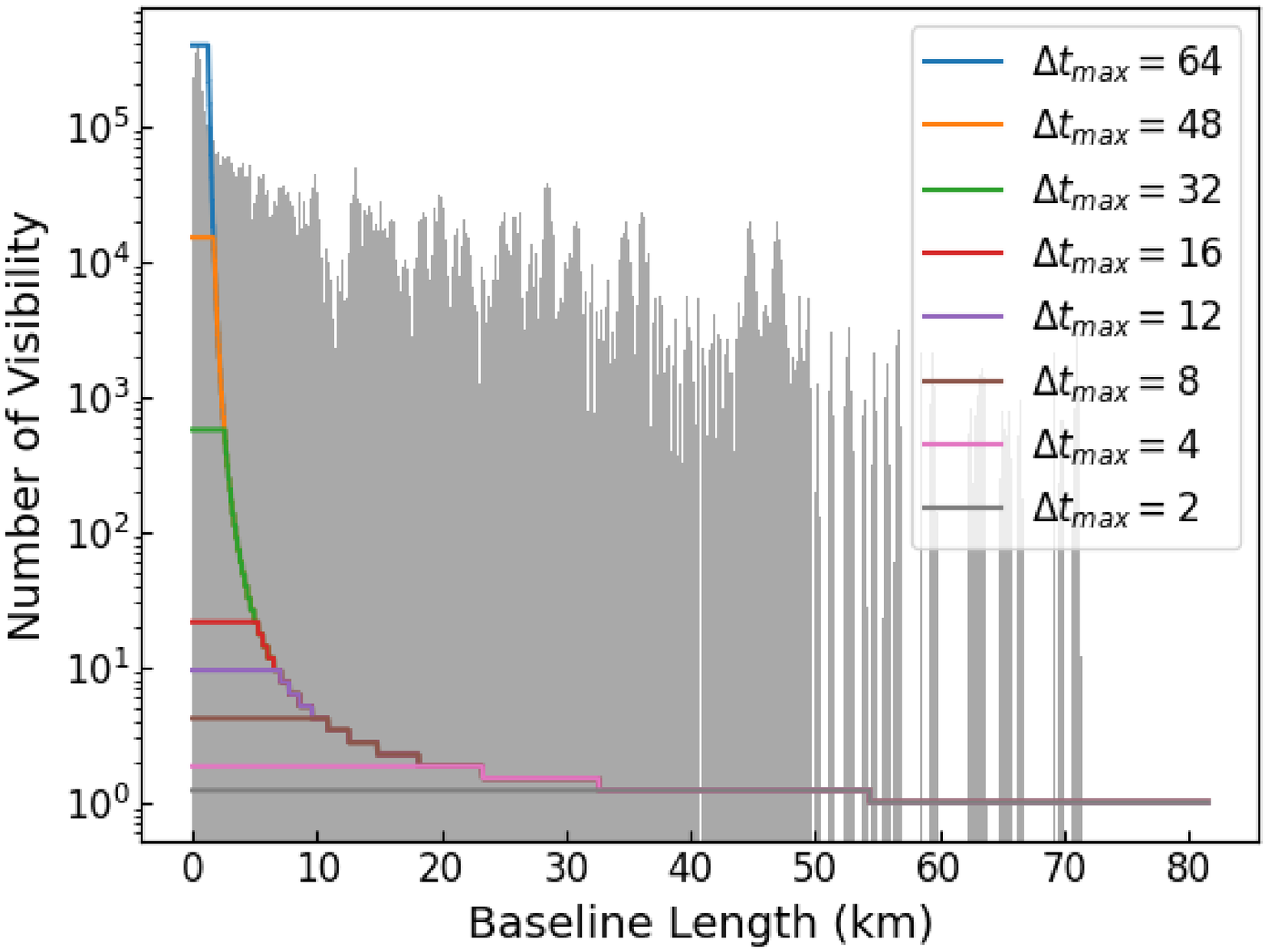}
        \caption{{The distribution of visibilities with baseline length, and the effect curves of the compression ratio.} }
        \label{ms2021-0369fig5}
    \end{minipage}%
    \hspace{.15in}
    \begin{minipage}[t]{0.47\textwidth}
        \centering
        \includegraphics[width=70mm]{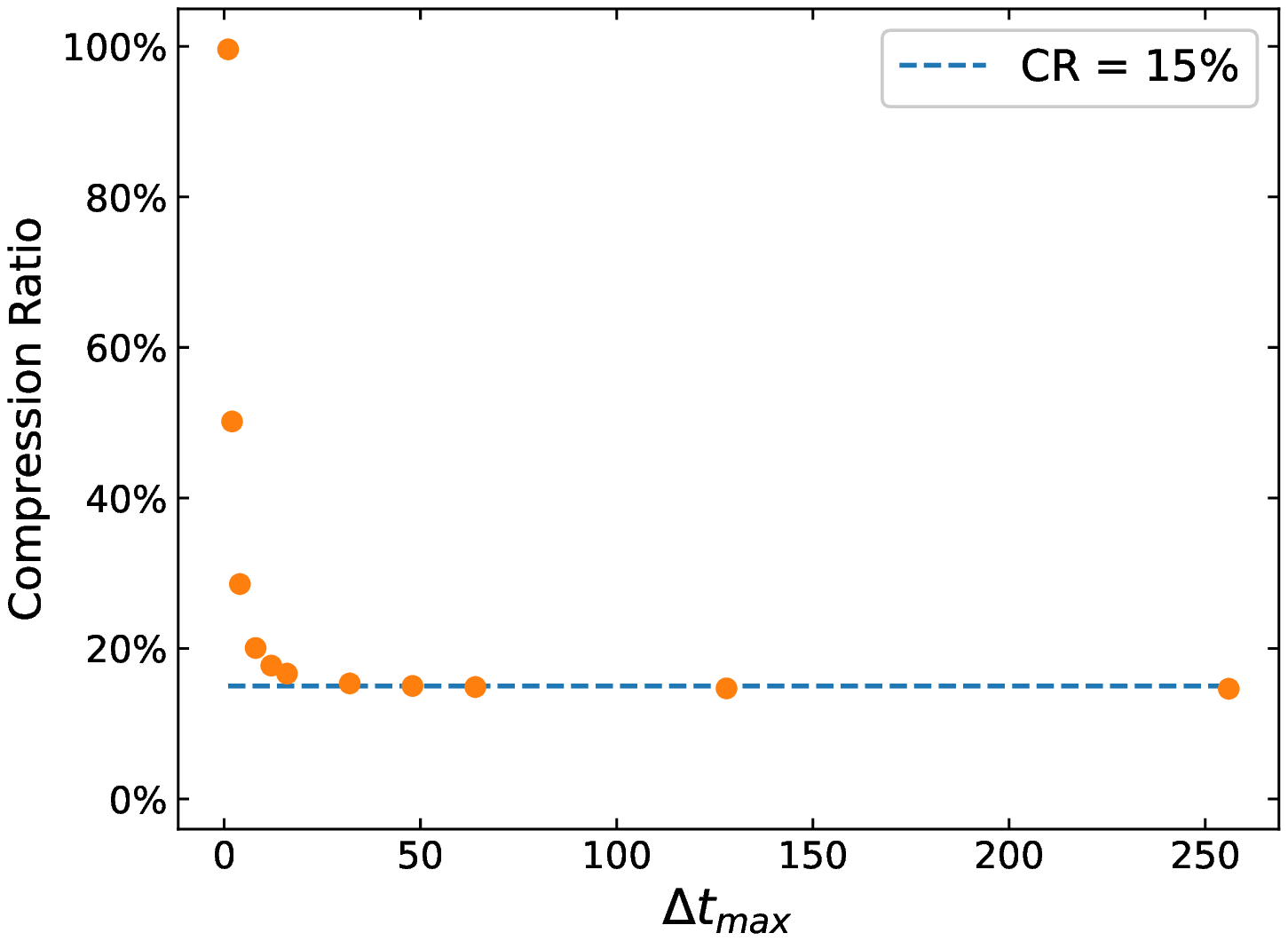}
        \caption{{The trend chart of the compression ratio with the maximum number of samples in averaging $\Delta t_{max}$.}}
        \label{ms2021-0369fig6}
    \end{minipage}%
\end{figure}

We set different upper limits ($\Delta t_{max}$), i.e.,  1, 2, 4, 8, 12, 16, 32, 48, 64, 128 and 256, to evaluate the compression ratio of the BDA. $\Delta t_{max}$ also means the maximum number of samples being averaged together on the shortest baselines.
These different upper limits lead to different compression ratios over the baseline length range, and some of these variation curves are shown in Figure \ref{ms2021-0369fig5}. Since the shorter baselines have larger data volumes, we want to average the amount of data over the shorter baselines as much as possible. We can also find that the increase of $\Delta t_{max}$ only further compresses the data on the shorter baselines, while the volume proportion of these data is decreasing in the total. 

The final result of the compression ratio with different $\Delta t_{max}$ is shown in Figure \ref{ms2021-0369fig6}. With the increasing value of $\Delta t_{max}$, the compression ratio reduces quickly and then becomes slow. Finally, a larger $\Delta t_{max}$ does not significantly improve the final compression ratio. It changes very little after the compression ratio reaches 15\%, where $\Delta t_{max} = 48$.
To avoid the more severe errors that may arise from a bigger $\Delta t_{max}$, it is worth considering using a $\Delta t_{max}$ less than 48 in subsequent studies.

\subsubsection{Imaging Quality Evaluation}

In addition to the compression ratio, the impact on the image quality after applying BDA is a significant issue. We defined the visibility data of the simulated observation as raw visibilities, the visibility data processed by the BDA as the averaged visibilities, and the final decompressed visibility data as recovered visibilities. We first decompressed the averaged visibilities by using a method similar to linear interpolation and then used the recovered visibilities for the subsequent imaging processing.

\begin{figure}[h]
    \begin{minipage}[t]{0.47\linewidth}
        \centering
        \includegraphics[width=7cm]{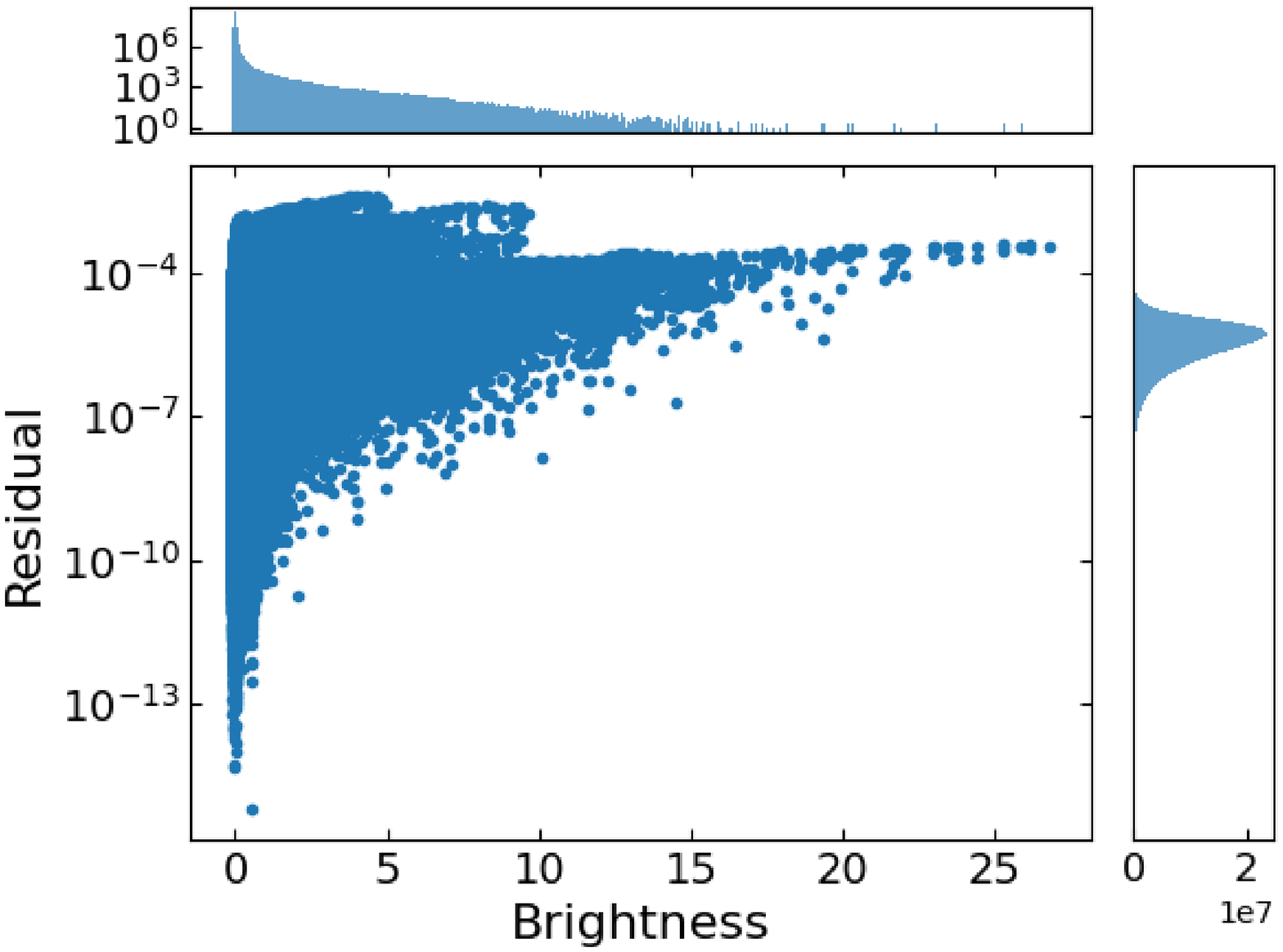}
    \end{minipage}%
    \begin{minipage}[t]{0.47\textwidth}
        \centering
        \includegraphics[width=7cm]{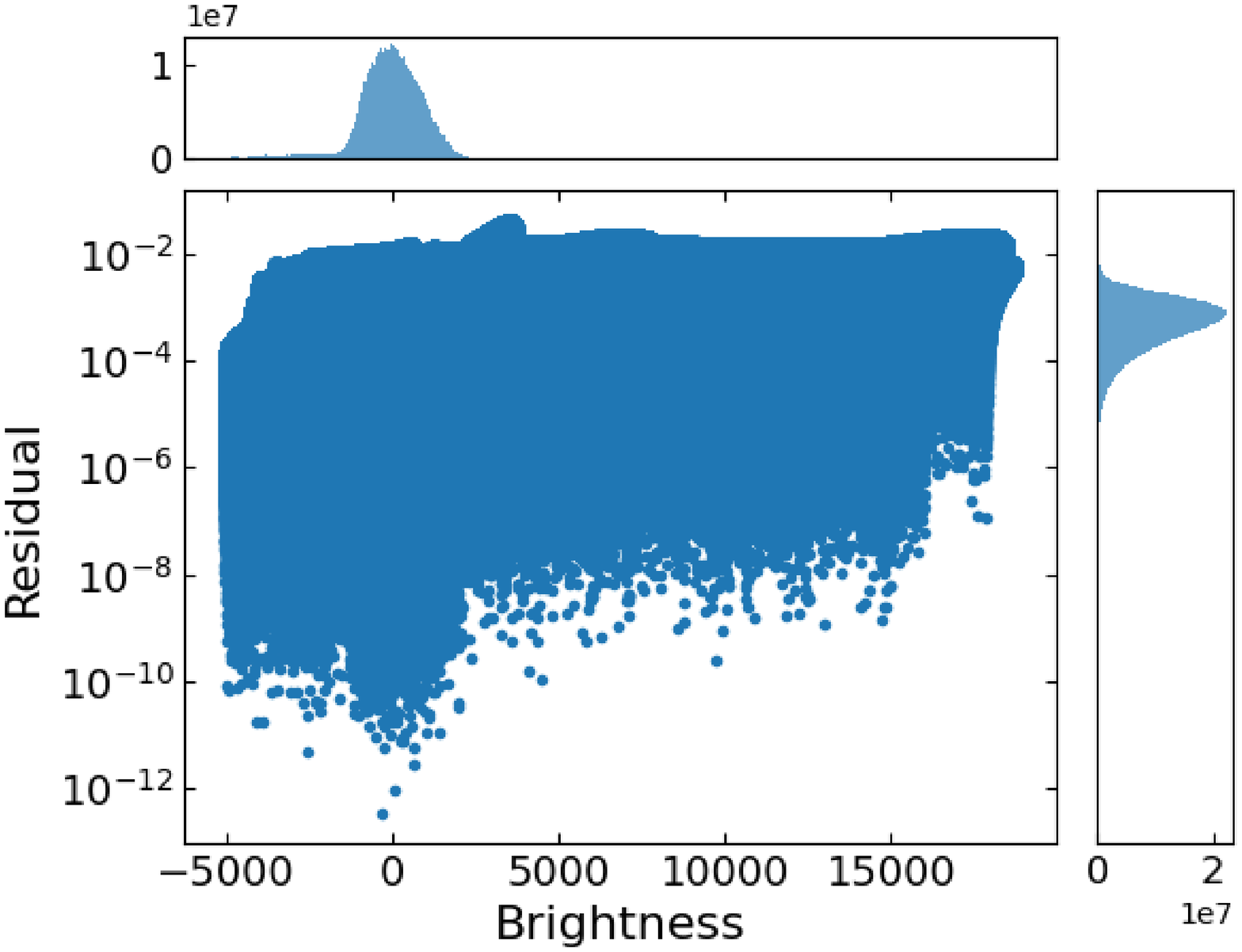}
    \end{minipage}%
    \centering
    \caption{
    The distribution of the absolute value of the residual with brightness, corresponding to the dirty image results of the GLEAM (left) and the M31 (right) as observed model images, and $\Delta t_{max}$ = 48.
    }
    \label{ms2021-0369fig7}
\end{figure}

To exclude the possible effects of different deconvolution methods on the imaging results, we used dirty images to analyze the imaging quality. 
Due to the difference between the recovered visibilities and the raw visibilities, the brightnesses in the dirty images are not the same. The deviations may be positive or negative relative to the dirty image of the raw visibilities. For convenience, the absolute value of the deviation is used here, and its distribution with the brightness is shown in Figure \ref{ms2021-0369fig7}, where $\Delta t_{max}$ of the recovered visibilities is equal to 48.

The brightness distribution of the dirty images is mainly concentrated around 0, and the lower limit of deviation increases with brightness, but the upper limit does not change excessively. At the same time, the maximum value of the residuals is small.

\begin{figure}[h]
    \begin{minipage}[t]{0.47\linewidth}
        \centering
        \includegraphics[width=66mm]{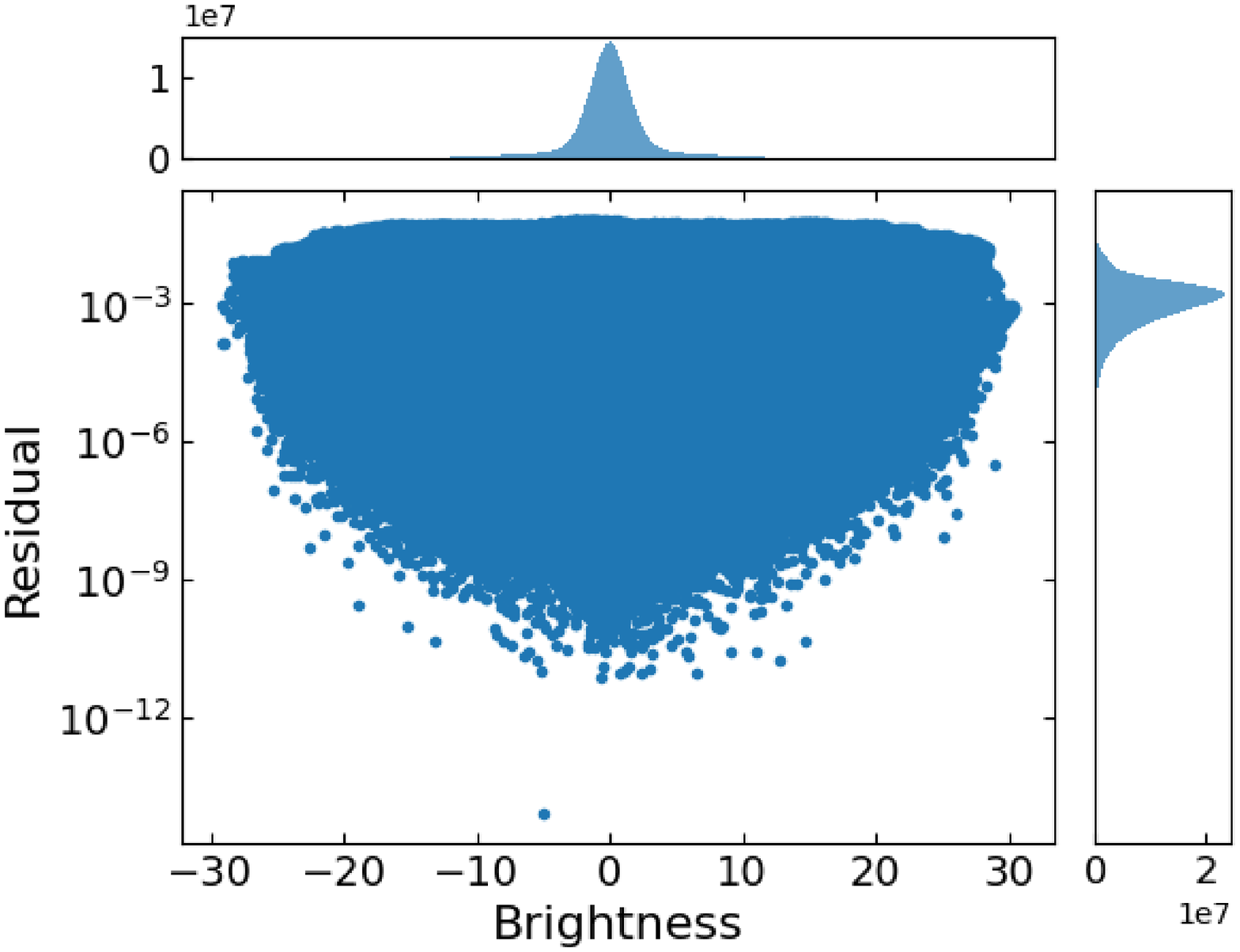}
        \caption{{The distribution of the absolute value of the residual with brightness for the pure noise model results, and $\Delta t_{max}$ = 48.}}
        \label{ms2021-0369fig8}
    \end{minipage}%
    \hspace{.15in}
    \begin{minipage}[t]{0.47\textwidth}
        \centering
        \includegraphics[width=69mm]{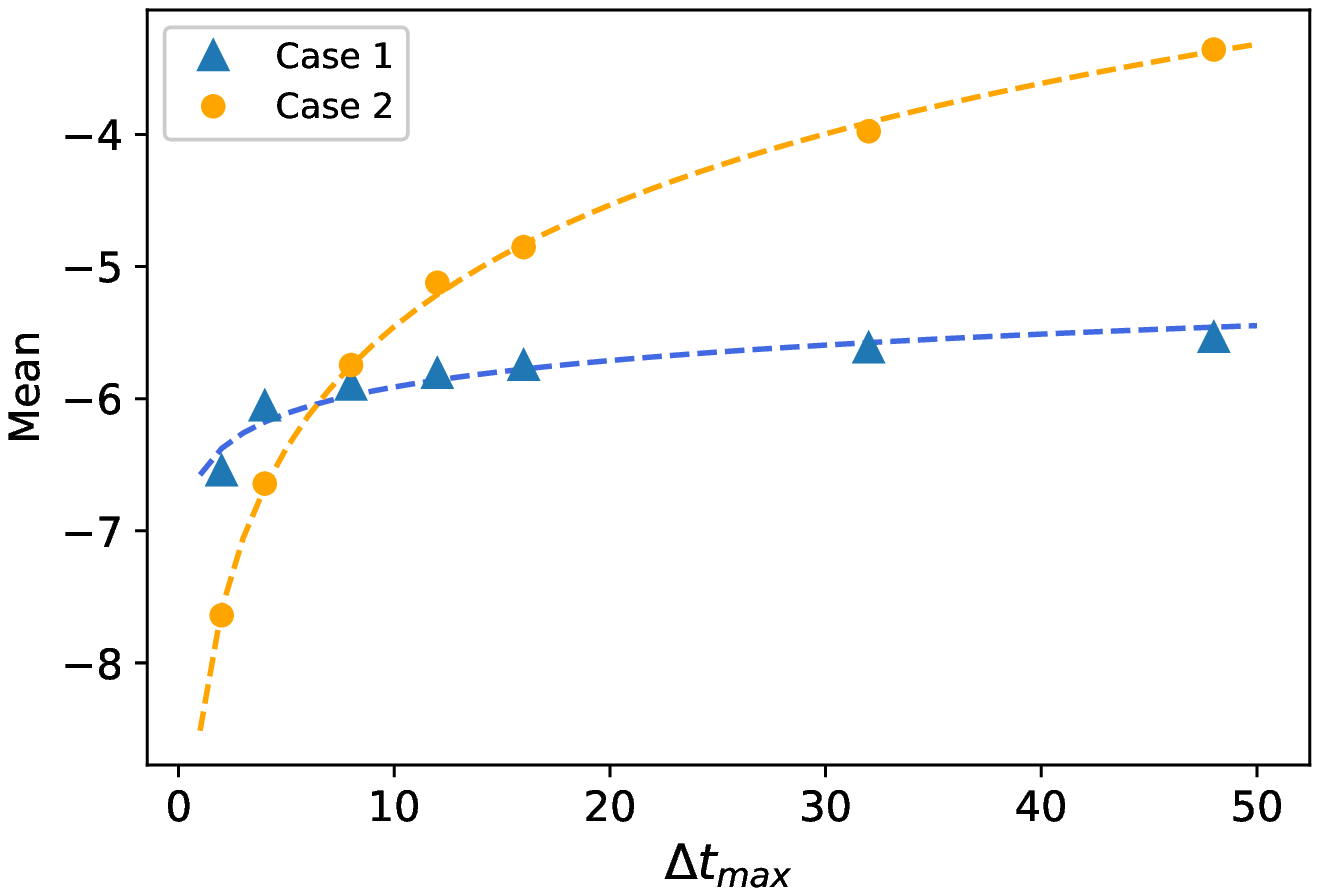}
        \caption{{The trend of the different mean values with $\Delta t_{max}$ in two groups.}}
        \label{ms2021-0369fig9}
    \end{minipage}%
\end{figure}

We used a pure noise image as a sky model for simulated observations and performed the same BDA processing. In generating this noisy model image, the same image size as the previous model was used, filled only with Gaussian noise with a mean of 0 and a standard deviation of 0.1.
Figure \ref{ms2021-0369fig8} shows the result of the dirty image when $\Delta t_{max}$ is equal to 48, where the maximum error is 0.0698. Moreover, when $\Delta t_{max}$ is equal to 2, the maximum error is 0.0162.
During this experiment, we tried to reduce the overall amplitude of the noisy model image by a certain ratio. The corresponding change in the dirty image is that both deviation and brightness values are reduced by the same ratio, while the contours of Figure~\ref{ms2021-0369fig8} do not change much.

The statistical distribution of the residuals shows a Gaussian-like distribution in the dirty image results when displayed in logarithmic form. The mean and standard deviation of the residuals in this form are given in Table \ref{ms2021-0369tab1}, where Case 1 refers to the results of dirty images for the GLEAM, and Case 2 refers to the one for the M31. 

The standard deviations of these two cases are approximately the same and do not change significantly with $\Delta t_{max}$. This indicates that the change in residual is more like an overall shift, while the mean is the distance of the shift. 

\begin{table}[h]
\begin{center}
    \caption[]{Mean and standard deviation of the logarithmic residual in two cases.}
    \label{ms2021-0369tab1}
    \begin{tabular}{cccccccc}
        \hline\noalign{\smallskip}
            & & \multicolumn{3}{c}{ Case 1 } & \multicolumn{3}{c}{ Case 2}\\   
        \cmidrule(r){3-5} \cmidrule(r){6-8}
            $\Delta t_{max}$  & CR & Mean & Std. & Max & Mean & Std. & Max \\ 
        \hline\noalign{\smallskip}
             2 & 50.16\% & -6.5449 & 0.5857 & -3.1033 & -7.6400 & 0.5309 & -5.4398 \\
             4 & 28.56\% & -6.0538 & 0.5570 & -2.6526 & -6.6431 & 0.5292 & -4.4348 \\
             8 & 20.06\% & -5.8964 & 0.5514 & -2.5298 & -5.7456 & 0.5250 & -3.6300 \\
            12 & 17.72\% & -5.8079 & 0.5485 & -2.4826 & -5.1231 & 0.5366 & -3.0849 \\
            16 & 16.66\% & -5.7466 & 0.5460 & -2.4582 & -4.8532 & 0.5250 & -2.7622 \\
            32 & 15.35\% & -5.6147 & 0.5401 & -2.4191 & -3.9763 & 0.5264 & -1.9117 \\
            48 & 15.02\% & -5.5350 & 0.5378 & -2.3999 & -3.3577 & 0.5440 & -1.3557 \\
        \noalign{\smallskip}\hline
    \end{tabular}
\end{center}
\end{table}

Figure \ref{ms2021-0369fig9} shows the relationship between the means and $\Delta t_{max}$ for these two cases, fitted with a logarithmic function for each. The same is that a small $\Delta t_{max}$ corresponds to a small imaging error, while case 2 has a larger variation range of the means than case 1.
This difference is probably due to the different characteristics of the amplitude intensity distribution of visibilities on the $uv$ plane in these two cases. Case 1 is relatively uniform, while case 2 is more concentrated in the low-frequency part. 

Considering both the compression ratio and dirty image quality, a small $\Delta t_{max}$ (e.g. $\Delta t_{max}$ = 12.) could meet the requirements of common use. 
We also found that further compression over short baselines is the cause of the relative error in the dirty images. A large $\Delta t_{max}$ implies large deviations in the recovered visibilities on short baselines.
In practice, it is difficult to invert a suitable $\Delta t_{max}$ from the imaging results, while choosing a small one is feasible and safe.

\subsection{The Processing Performance}

The processing performance of BDA is a fundamental metric.
A series of tests were performed on a Centos 7 server equipped with 32 processors (Intel Xeon Gold 6226R), 2.9-GHz core frequency, and 1024 GB of RAM. 
The version of RASCIL used to obtain the simulation data in the tests was v.0.1.11.
Using a 12-minute single-channel simulation data of SKA1-LOW with a data volume of 12.5 GB, we profile the BDA module optimized by the Numba, pure Python code, and Pandas, respectively. 
The performance results are presented in Table \ref{ms2021-0369tab2}. The BDA module implemented using the Numba has the best performance.  

\begin{table}[h]
\begin{center}
    \caption[]{The performance of the BDA implementation at different data volumes.}
    \label{ms2021-0369tab2}
    \begin{tabular}{cccc}
        \hline\noalign{\smallskip}
            Data volume & Numba & Pandas & Pure Python \\
            (GB) & (seconds) & (seconds) & (seconds) \\
        \hline\noalign{\smallskip}
            1.56 & 10.75 & 13.69 & 400.01 \\
            3.13 & 17.48 & 23.78 & 697.62 \\
            6.25 & 29.02 & 44.21 & 1413.21 \\
            12.50 & 58.79 & 86.18 & 2941.89 \\
            25.00 & 105.59 & 171.38 & 6523.35 \\
            50.00 & 243.12 & 367.48 & 16178.32 \\
        \noalign{\smallskip}\hline
    \end{tabular}
\end{center}
\end{table}

\begin{figure}[h]
    \centering
    \includegraphics[width=9 cm]{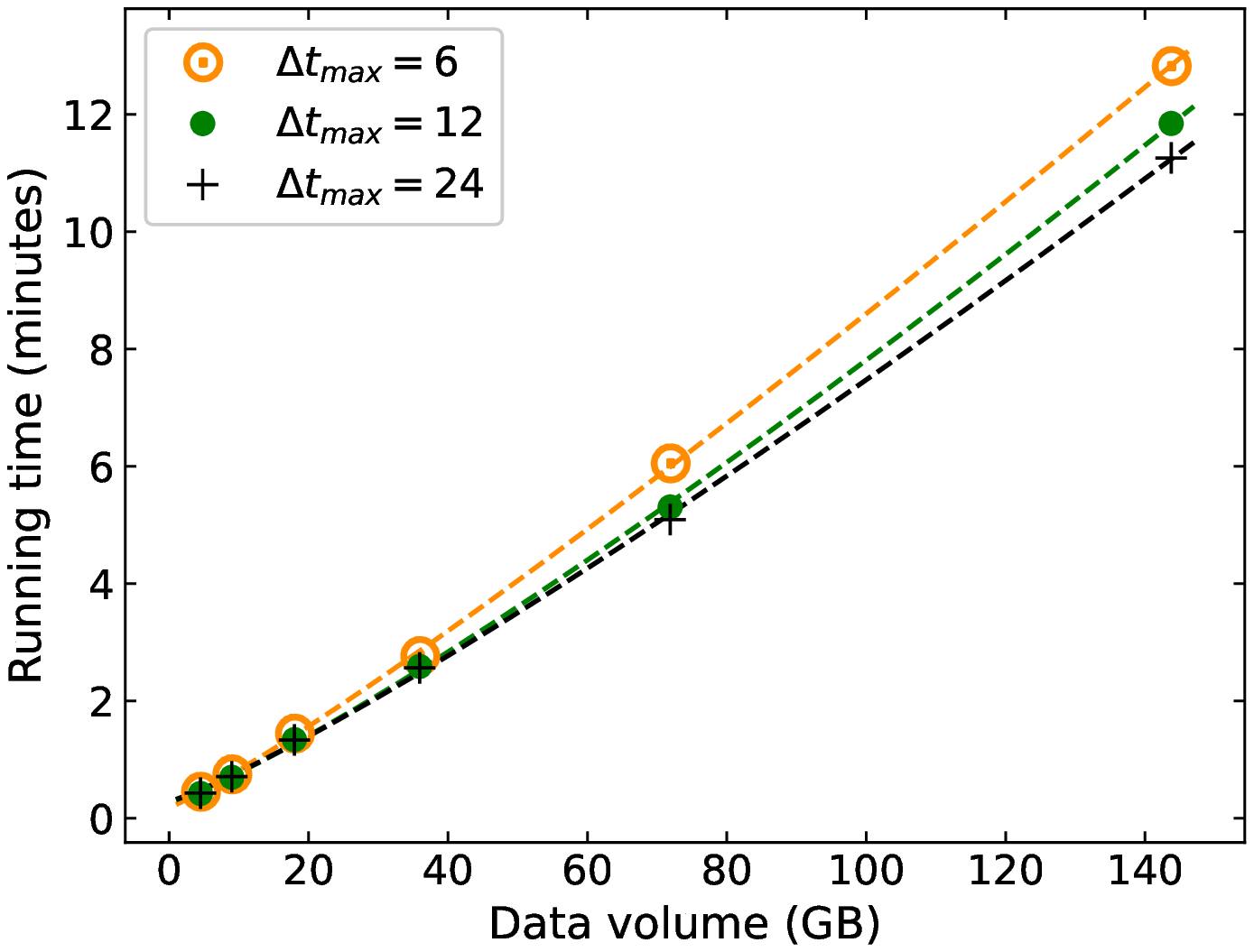}
    \caption{The performance of the BDA implementation optimized with Numba at different data volumes.}
    \label{ms2021-0369fig10}
\end{figure}

Therefore, we further tested the processing performance of the Numba optimized code with a series of simulated data of different observation times and four channels. The maximum observation time was 48 minutes, and the data volume obtained was 143.75 GB. During the process of BDA, $\Delta t_{max}$ was set to 6, 12, and 24 respectively, and $\Delta f_{max}$ was 1. The performance result is shown in Figure \ref{ms2021-0369fig10}, fitted with the exponential function respectively.

As the results are shown in Figure \ref{ms2021-0369fig10}, the time consumption of BDA processing is essentially linear with the amount of data to be processed. The processing speed of BDA in the case of a single process is about 13 GB per minute. Further improvements are expected under parallel computing conditions. This speed is acceptable for SRCs data pre-processing. In addition, the processing speed has little to do with the amount of $\Delta t_{max}$.

\section{Discussions}

Experimental results showed a significant decrease in the capacity of the visibility data with different $\Delta t_{max}$, which indicates that BDA is very valuable for SKA data processing, especially for the construction of subsequent SRCs.
The annual storage capacity of SRCs will be at least 5 petabytes (PB) per year in the beginning, which will increase to at least 1.7 Exabytes by 2028, with at least 700 PB online (\citealt{Bolton+2019}). BDA can compress at least 50\% of the short baseline visibility data, which is valuable for reducing the cost of SRCs construction.

This study also gave the variation of the imaging quality at different $\Delta t_{max}$. The experimental results provided an essential reference for SKA1-LOW to carry out EoR and cosmic dawn. The MWA data were averaged (\citealt{Mitchell+2008}), but no specific details were given.

\section{Conclusions}

We implemented a new BDA module based on RASCIL, and this implementation is created by designing functions using Numba. It has not introduced excessive memory usage in the tests and completes the computation tasks faster than the other modules. The speed of a single process is around 13 GB per minute. It also performed well during the BDA processing of the simulation data. 

Through the simulation of observing the GLEAM and M31 model images with SKA1-LOW, we evaluate the performance of the BDA. According to the subsequent analysis, the error due to BDA increases with the maximum upper limit of the averaging interval on the short baselines. In contrast, the compression ratio does not improve all the time, and the reduction in data volume remains at a maximum ratio of approximately 85\%.
The smaller upper limit is sufficient for the compression ratio, and the imaging error is reasonable. Overall, the BDA technology will have applications in the face of massive SKA observation data processing. The BDA can effectively reduce the storage space of visibility data, as it is also valuable for the future construction and application of SRCs.

\begin{acknowledgements}
This work is supported by the National SKA Program of China (2020SKA0110300), the Joint Research Fund in Astronomy (U1831204, U1931141) under cooperative agreement between the National Natural Science Foundation of China (NSFC) and the Chinese Academy of Sciences (CAS), the Funds for International Cooperation and Exchange of the National Natural Science Foundation of China (11961141001), the National Natural Science Foundation of China (No.11903009). 
This work is also supported by Astronomical Big Data Joint Research Center, co-founded by National Astronomical Observatories, Chinese Academy of Sciences and Alibaba Cloud.

We do appreciate the anonymous referee for valuable and helpful comments and suggestions. 

\end{acknowledgements}

\bibliographystyle{raa}
\bibliography{ms2021-0369bibtex}

\end{document}